\documentclass[prl,twocolumn,floatfix]{revtex4}

\usepackage{amsfonts}
\usepackage{amsmath}
\usepackage{verbatim}
\usepackage[dvips]{graphicx}
\usepackage{subfigure}

\def\dd{\textrm{d}}
\def\Bf{\boldsymbol}

\def\rv{\Bf{r}}

\def\kv{\Bf{k}}

\def\qv{\Bf{q}}
\def\Qv{\Bf{Q}}
\def\Gv{\Bf{G}}
\def\nv{\Bf{n}}

\begin{document}

\title{Finite Momentum Pairing Instability of Band-Insulators With Multiple Bands}
\author{Predrag Nikoli\'c$^1$, A.A. Burkov$^2$, Arun Paramekanti$^3$}
\affiliation{$^1$Department of Physics, Rice University, Houston, TX 77005, USA}
\affiliation{$^2$Department of Physics and Astronomy, University of Waterloo, Waterloo,  Ontario N2L 3G1, Canada}
\affiliation{$^3$Department of Physics, University of Toronto, Toronto, Ontario M5S 1A7, Canada}
\date{\today}

\begin{abstract}
We  show, based on microscopic models, that fermionic band insulators with multiple bands and strong
interband attraction are
generically unstable towards nonzero momentum Cooper pairing leading to a
pair density wave (PDW) superfluid state. Our first model considers a band insulating state of fermionic atoms
in a three-dimensional cubic optical lattice. We show that this insulator
is unstable towards an incommensurate PDW in the vicinity of a Feshbach resonance.
Our second model is a two-band tight binding model relevant to electrons in solids; we show that the
insulating state of this
model has a PDW instability analogous to the
exciton
condensation instability in indirect bandgap semiconductors. We discuss relevant experimental signatures of the PDW
state.

\end{abstract}

\maketitle

{\it\bf Introduction. ---} The theme of coexisting or competing order parameters is common to several strongly correlated systems including high temperature cuprate \cite{cuprates} and pnictide \cite{Norman} superconductors. Most notably,
several cuprate materials exhibit
stripes or checkerboard patterns of spin and charge modulations that coexist with superconductivity
\cite{Tranquada, Ali, McElroy, Hanaguri, Fang}. Motivated by the observation \cite{tranquada07} of quasi-two-dimensional superconductivity  coexisting with stripe order in the layered superconductor La$_{1.875}$Ba$_{0.125}$CuO$_4$ (LBCO), Berg {\it et al.}~\cite{Kivelson} have proposed, on phenomenological grounds, that a distinct state of matter, named a `pair density wave' (PDW), is realized in this material. In its simplest avatar, the PDW state results from condensing singlet Cooper pairs with nonzero center-of-mass momenta $\pm\Qv$ and is accompanied by an induced charge density modulation at momenta $\pm 2 \Qv$. It is thus similar to the Fulde-Ferrell-Larkin-Ovchinnikov (FFLO) state of magnetized superfluids \cite{FFLO} except that the PDW does not require
a spin population imbalance \cite{Leo}.
In contrast to earlier proposals for a Cooper pair insulator \cite{Zhang04,Tesanovic04} in LBCO, the PDW state is a supersolid, in that it has coexisting superfluid and density orders, which break lattice symmetries. However, it is very different from the supersolid state proposed to exist in $^4\textrm{He}$ \cite{KimChan,andreev}, or the supersolids realized in simple lattice models \cite{Melko05}, since the superfluid order parameter in the PDW state has no uniform Fourier component. The bosonic analog of the PDW occurs in lattice models in which the boson kinetic energy is `frustrated' so that
bosons condense into multiple modes with nonzero momenta \cite{Burkov06}.

The main contribution of this work is to show there are simple microscopic models of fermions, relevant
to cold atomic gases and solid state materials,
which support a PDW ground state. Our work goes beyond earlier Landau theory descriptions and
Josephson junction models
of the PDW state \cite{Kivelson}.
Our first example is a one-channel model of fermionic atoms near a Feshbach
resonance \cite{Bloch,unit}  confined to a cubic optical lattice.
It has been demonstrated recently~\cite{Ketterle2} that this system shows a
superfluid to band-insulator transition \cite{optlat2,optlat1,optlat3}
when the lattice depth is varied at a commensurate density of two atoms per lattice site.
Here we show, via a more careful study,
that a PDW state is expected to intervene between the uniform superfluid and
the band insulator. Our second example is a two-band tight binding
model where an appropriate choice of local
attractive interactions between the fermions leads to the PDW instability of a band
insulator. We discuss direct and indirect experimental signatures
of PDW order in these systems as well as the experimental feasibility of achieving such states.

The key physics which leads to the emergence of the PDW state in both these models is
the presence of multiple bands and
the dominance of interband Cooper pairing. In the cold atom model, we present
arguments to show that, in contrast to intraband pairing,
the phase space for interband pairing is expanded at nonzero pairing momenta, which
stabilizes an incommensurate PDW state.
In the two-band
tight binding model, the reason for the occurrence of the PDW state is that the lowest
energy momentum points in each band differ by a nonzero wavevector $\Qv$,
which leads to a large Cooper pair
susceptibility at this wavevector.
For this model, we present the mean field phase diagram and show that
the PDW instability is closely related to
the Halperin-Rice exciton condensation instability in indirect bandgap semiconductors~\cite{HalperinRice},
and some models of Iron-pnictides~\cite{Zlatko}.

{\it\bf Cold Atoms in an Optical Lattice. ---} We describe fermionic atoms with attractive interactions in a periodic potential \cite{optlat2,optlat1,optlat3} using the Hamiltonian ($\hbar=1$):
\begin{equation}\label{Sfull}
H = \int \dd^3 r \Biggl\lbrack c_{\sigma}^{\dagger}
  \left( -\frac{\boldsymbol{\nabla}^2}{2m}-\mu+V_{\rv} \right) c_{\sigma}^{\phantom{\dagger}}
  - U c_{\uparrow}^{\dagger} c_{\downarrow}^{\dagger}
         c_{\downarrow}^{\phantom{\dagger}} c_{\uparrow}^{\phantom{\dagger}}
   \Biggr\rbrack.
\end{equation}
Owing to universality in the unitarity regime, this simple theory provides a faithful description of fermionic cold atoms tuned near a broad Feshbach resonance \cite{unitary}. We will study this model using mean-field theory which is known to
be a reasonable approximation near unitarity for the qualitative points we wish to make.
Fluctuations can be treated systematically using, for example, large-$N$ expansions \cite{optlat1,unitary,rvs},
but we will not pursue this here.

We work with a simple cubic lattice potential:
\begin{equation}\label{Pot}
V_{\rv} = V \left\lbrack \cos\left(\frac{2\pi x}{a_L}\right) +
  \cos\left(\frac{2\pi y}{a_L}\right) + \cos\left(\frac{2\pi z}{a_L}\right)
  \right\rbrack \ ,
\end{equation}
where $a_L$ is lattice spacing. The quantum numbers of single-particle Bloch eigenstates in this potential are crystal wavevector $\kv=(k_x,k_y,k_z)$ inside the first Brillouin zone (BZ) $-\pi/a_L\le k_x,k_y,k_z < \pi/a_L$, and band index $\nv=(n_x,n_y,n_z)$. We label the Bloch wavefunctions by $\psi_{\nv\kv}(\rv)$ and the corresponding energies by $\epsilon_{\nv\kv}$.

Near unitarity, the cutoff-dependent contact interaction parameter $U$ is related to the scattering length $a$:
\begin{equation}
\label{scatlength}
\frac{1}{U} = - \frac{m}{4 \pi a} +   \sum_{\nv}\int\frac{\dd^3 k}{(2\pi)^3} \frac{1}{2 \epsilon_{\nv\kv}}\Bigl\vert_{V=0}.
\end{equation}
Band-index cutoff, discussed below, is implicit in (\ref{scatlength}).

A $T=0$ superfluid-insulator transition for an even number of fermions per site occurs at a critical value of the lattice amplitude $V$, which is a universal function of $a_L/a$ and the fermion density \cite{optlat1,optlat2,optlat3}. Starting from a band-insulating state, the onset of pairing in the mean-field approximation can be extracted from the inverse static pairing susceptibility matrix:
\begin{eqnarray}\label{MatrixK}
&& \Pi_{\Gv\qv;\Gv'\qv'} =
    \sum_{\nv_1\nv_2} \int \frac{\dd^3 k_1}{(2\pi)^3} \frac{\dd^3 k_2}{(2\pi)^3}
    \frac{f\left(\xi_{\nv_1 \kv_1}\right) - f\left(-\xi_{\nv_2 \kv_2}\right)}
         {\xi_{\nv_1 \kv_1} + \xi_{\nv_2 \kv_2}} \nonumber \\
&& ~~ \times \Gamma_{\nv_1 \kv_1 ; \nv_2 \kv_2}^{\Gv \qv *} \Gamma_{\nv_1 \kv_1 ; \nv_2 \kv_2}^{\Gv' \qv'} +
             \frac{(2\pi)^3}{U} \delta(\qv-\qv')\delta_{\Gv \Gv'},
\end{eqnarray}
where $\qv$ are first BZ wavevectors, $\Gv$ are reciprocal lattice vectors,  $\xi_{\nv \kv}= \epsilon_{\nv \kv} - \mu$, $f(\xi)$ is Fermi-Dirac distribution function and $\Gamma$ are vertex functions:
\begin{equation}\label{Ver}
\Gamma_{\nv_1 \kv_1 ; \nv_2 \kv_2}^{\Gv \qv} = \int\dd^3 r \Phi_{\Gv \qv}^*(\rv)
  \psi_{\nv_1 \kv_1}^{\phantom{*}}(\rv) \psi_{\nv_2 \kv_2}^{\phantom{*}}(\rv).
\end{equation}
Since crystal momentum is conserved, $\Pi_{\Gv\qv;\Gv'\qv'} = \Pi_{\Gv\Gv'}(\qv)\times (2\pi)^3 \delta(\qv-\qv')$.
We will use the plane wave representation for $\Pi_{\Gv\Gv'}(\qv)$, corresponding to the pair wavefunctions $\Phi_{\Gv \qv}(\rv) = e^{i(\qv+\Gv)\rv}$. All eigenvalues of the matrix $\Pi_{\Gv\Gv'}$ are positive in the band insulating state. When the lowest eigenvalue $\Pi(\qv)$ becomes negative at some wavevector $\qv=\Qv$, the insulating state becomes unstable to a superfluid of fermion pairs condensing at momentum $\Qv$, which is a PDW state~\cite{Kivelson} if $\Qv\neq 0$.

\begin{figure}
\includegraphics[height=2.2in]{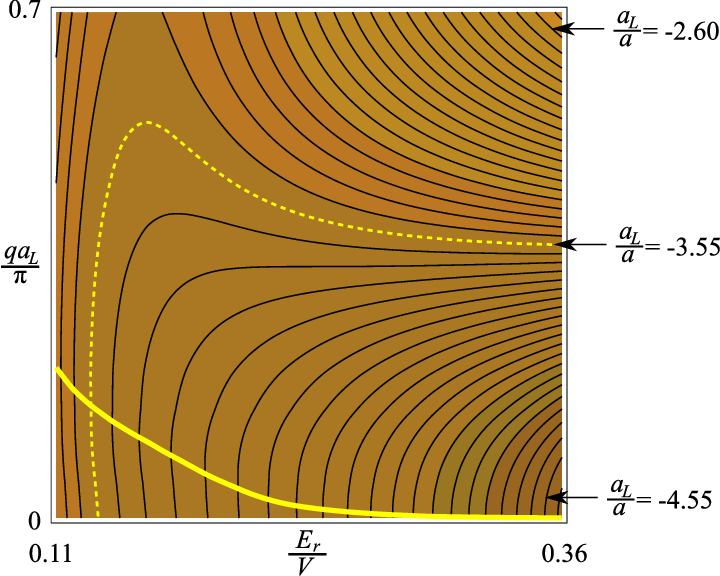}
\caption{\label{Qcritical} Critical curves of the inverse Cooper pair susceptibility
$\Pi(\qv)$, at which its lowest eigenvalue changes sign,
for $\qv=(q,q,q)$ in the band-insulator with two atoms per site and
various scattering lengths. Bright solid line shows the PDW wavevector $Q$ at the transition
as a function of inverse lattice depth ($E_r=\pi^2/4ma_L^2$ is molecular recoil energy).}
\vskip -0.2in
\end{figure}

Normally one would expect the pair condensation to occur at $\Qv=0$.  This is certainly true in any single-band model of lattice fermions. However, as we demonstrate below, interband pairing in multi-band models can give rise to pairing
instability at a finite $\Qv$. Figure \ref{Qcritical} shows the critical curves (for two fermions per site at $T=0$)
at which the lowest eigenvalue of $\Pi(\qv)$ changes
sign for a given scattering length $a$, signalling an instability of the band insulator \cite{calculation}.
Coming from the deep lattice limit, $E_r/V \ll 1$, it is clear that the first instability one encounters (corresponding to
the leftmost point on each contour) occurs at nonzero momentum $\qv=\Qv$ for a wide range of scattering lengths.
The smooth evolution of $|\Qv|$ with the lattice depth in the deep BCS limit indicates that the formed PDW state is incommensurate. As the pairing interactions become stronger, in the BEC regime, $|\Qv|$ grows and possibly eventually saturates at the BZ edge making the PDW commensurate although we could not explore this regime
numerically. Notably, sometimes a superfluid at large $\qv$ (which can be imposed by
a superflow) can be destabilized by both increasing and \emph{decreasing} $V$ (e.g., dashed contour in Fig.~1).
The latter illustrates that interband pairing is responsible for superfluidity at finite $\qv$, which can
be expected to weaken with decreasing $V$.
Without knowing the quartic terms in the Landau theory, we cannot rule out time-reversal symmetry breaking,
but such a calculation is prohibitively hard.
We next provide simple arguments to show why our multiband system can favor a PDW instability.

\begin{figure}
\subfigure[{}]{\includegraphics[height=1.3in]{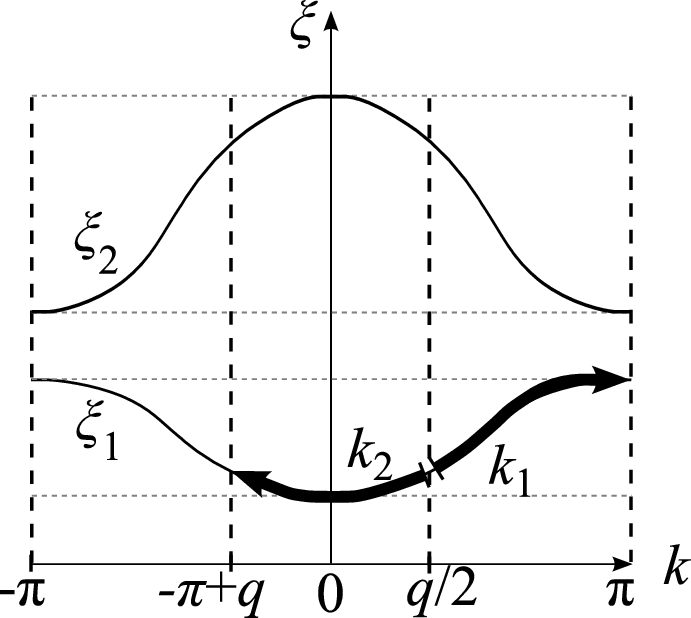}}
\subfigure[{}]{\includegraphics[height=1.3in]{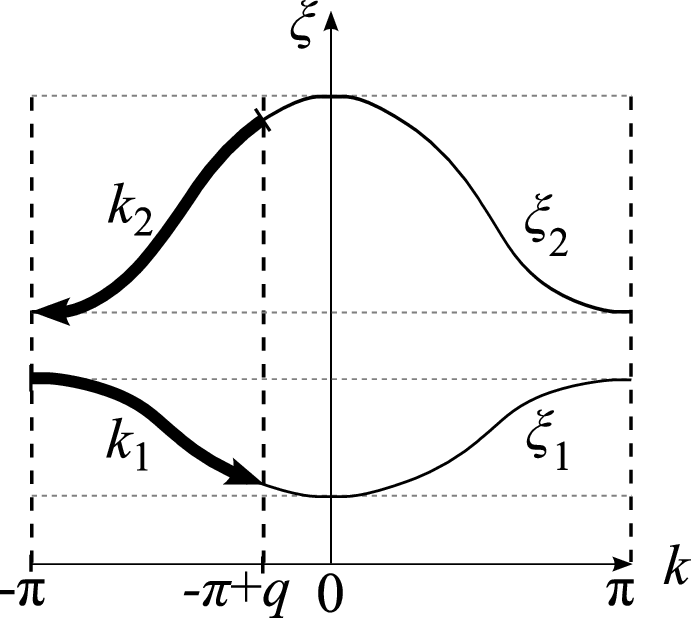}}
\caption{\label{Pairing}Pairing of two fermions with crystal momenta $k_1$ and $k_2$. Intraband pairing in (a) occurs when $k_1+k_2-q=0$. Interband pairing in (b) occurs when $k_1+k_2-q=-2\pi$ (assuming $q>0$). Thick arrows show the trajectories of $k_1$ and $k_2$ dictated by momentum conservation.}
\end{figure}

The \emph{incommensurate} PDW owes its existence to phase-space restrictions for interband pairing. Consider two bands along some momentum direction in the first BZ, separated by an indirect `gap' (which may be filled by other bands). Let us describe them using a one-dimensional toy model with $a_L=1$ as in Fig. \ref{Pairing}. Since momentum is conserved only modulo reciprocal lattice vectors $G$, we can rewrite the vertex functions ~(\ref{Ver}) as:
\begin{equation}\label{Ver2}
\Gamma_{n_1 k_1 ; n_2 k_2}^{G q} = \sum_{G'} A_{n_1 n_2}^{G q}(G') \times 2\pi\delta(k_1+k_2-q+G') \ ,
\end{equation}
where the coefficients $A_{n_1 n_2}^{G q}(G')$ depend on details of the band-structure. All of these coefficients for any fixed $(n_1, n_2, G)$ must gradually vanish in the $V \to 0$ limit, except one (at a particular value of $G'$) which approaches unity. For example, pairing into a plane-wave superfluid at $q\in\textrm{first BZ}$ is given by $A_{11}^q(G) \approx \delta_{G,0}$ (intraband) and $A_{12}^q(G) = A_{21}^q(G) \approx \delta_{G,2\pi\textrm{sgn}(q)}$ (interband) for small $V$. We simplify the following discussion by focusing only on this pairing channel which reduces the inverse pairing susceptibility matrix $\Pi_{GG'}(q)$ to a scalar $\Pi(q)$. Corrections due to condensate harmonics at larger reciprocal lattice vectors are negligible in the small $V$ limit. 

Using (\ref{Ver2}) we find that the main contribution to intraband pairing for small $V$ at $T=0$ comes from
\begin{equation}\label{IntraBand}
\Pi^{(1,1)}(q) \approx -\frac{1}{2\pi} \int \dd k_1 \dd k_2
   \frac{\delta(k_1+k_2-q)}{\xi_{1,k_1} + \xi_{1,k_2}}, \nonumber
\end{equation}
which is illustrated in Fig. \ref{Pairing}(a). Since $k_1$ and $k_2$ are restricted to the first BZ, the number of states available for intraband pairing \emph{decreases} with $q$. Consequently, the magnitude
of $\Pi^{(1,1)}(q)$ decreases with $q$ and thus purely intraband pairing would occur at $q=0$. The dominant interband contribution
\begin{equation}\label{InterBand}
\Pi^{(1,2)}(q) \approx -\frac{1}{2\pi} \int \dd k_1 \dd k_2
   \frac{\delta(k_1+k_2+2\pi\textrm{sgn}(q)-q)}{\xi_{1,k_1} + \xi_{2,k_2}}, \nonumber
\end{equation}
illustrated in Fig. \ref{Pairing}(b) has the opposite behavior because the number of states available for interband pairing \emph{increases} with $q$. Therefore, interband processes alone would prefer pairs to condense at a BZ edge. 

It is important to note that $\Pi(q) \sim |q|$ for $q\to 0$ due to the boundaries of momentum integrals in all $\Pi^{(n_1 n_2)}(q)$, as can be seen from Fig. \ref{Pairing}. Only for $V=0$ and in the tight-binding limit do these linear contributions cancel out, leading to  $\Pi(q)\sim q^2$. The initially negative slope of $\Pi(q)$ leads to a local minimum at $q\neq 0$. The location of this minimum is determined by the relative strengths of interband and intraband contributions, so that in principle it can be anywhere in the BZ, making the PDW generically incommensurate.

A linear $\Pi(\qv)$ for $\qv\to 0$ is incompatible with a uniform superfluid instability. Since phase-space restrictions for pairing in the presence of a periodic potential generally result in a linear $\Pi(\qv)$, we argue that a PDW supersolid \emph{always} preempts an ordinary superfluid instability of the band insulator. This is consistent with our numerical findings. Note that fluctuations beyond the mean-field approximation {\it cannot} destroy the PDW instability.

{\it\bf Two-band tight binding model. ---}
Let us next turn to a tight-binding model which is of interest for
fermions in deep optical lattices or for solid
state materials. We consider a multiband fermion Hamiltonian
\begin{eqnarray}
H \!\!\!& = &\!\!\!
  - \!\!\sum_{\langle i,j\rangle n \sigma} t^{\vphantom\dagger}_n \left( c_{i n\sigma}^{\dagger}
  c_{j n \sigma}^{\vphantom\dagger} \!+\! h.c. \right)
  \!\!+\!\! \sum_{i n \sigma}
  (\gamma^{\vphantom\dagger}_n\!-\!\mu)
  c_{i n\sigma}^{\dagger} c_{i n \sigma}^{\vphantom\dagger}
  \nonumber \\
\!\!\!&- &\!\! U \!\!\!\sum_{i,n_1,n_2,\ell_1,\ell_2} (
\lambda^{\vphantom\dagger}_{n_1 n_2} c^\dagger_{i n_1 \uparrow} c^\dagger_{i n_2 \downarrow})
(\lambda^{\vphantom\dagger}_{\ell_1 \ell_2}
c^{\vphantom\dagger}_{i \ell_2 \uparrow}
c^{\vphantom\dagger}_{i \ell_1 \downarrow}),
\end{eqnarray}
where the fermions have a band-index $n$ and spin $\sigma$. The single particle
dispersion is governed by hopping amplitudes $t_n$ and site energies $\gamma_n$.
We include attractive band-dependent
interactions parameterized by a strength $U$ and couplings $\lambda_{n_1 n_2}$.
For simplicity, we focus here on a two-dimensional two-band model and restrict ourselves to the
case where $\lambda_{11}=\lambda_{22} \equiv \cos\theta$, and $\lambda_{12}=-\lambda_{21}
 \equiv \sin\theta$, with $0\! \leq \!\theta\! \leq \! \pi/2$. With this parametrization, the overall pairing
 strength is controlled by $U$, while tuning the angle  $\theta$ takes us from pure intraband
 pairing ($\theta=0$) to pure interband pairing ($\theta=\pi/2$).

  \begin{figure}[t]
\includegraphics[width=2.2in]{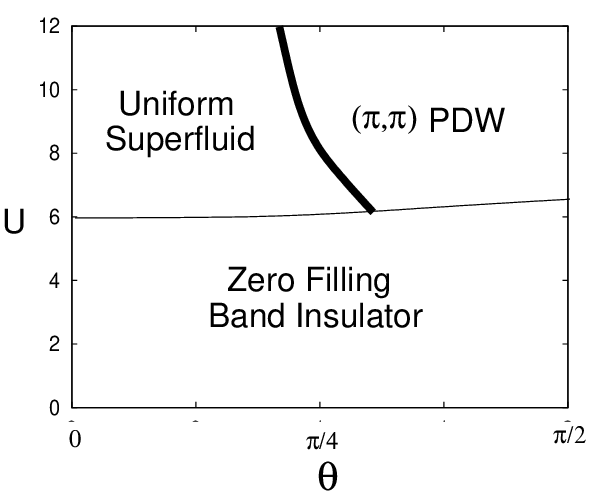}
\caption{\label{fig:pdw} Mean field phase diagram of the two-band tight-binding
Hamiltonian with pairing strength $U$ and
a parameter $\theta$ which tunes the interaction from pure intraband pairing ($\theta\!=\!0$)
to pure interband pairing ($\theta\!=\!\pi/2$) (see text for details).
We choose $t_1\!=\!1$, $t_2\!=\!-1$,$\gamma_1\!=\!0$, $\gamma_2\!=\!4$, and $\mu\!=\!-4.5$.
Thin (thick) lines indicate second (first) order transitions.}
\end{figure}

Fig.~\ref{fig:pdw} shows the mean field phase diagram of this model for a specific choice of dispersion and
chemical potential at which the noninteracting state is a zero-filling band insulator. We find that this band
insulator can undergo continuous transitions into either a uniform superfluid or a PDW state depending on
whether intraband or interband interactions dominate.
To understand this phase diagram, we compute the
inverse Cooper pair susceptibility of the band insulator,
\begin{equation}
\Pi({\qv}) = \frac{1}{U} + N_s^{-1} \sum_{\kv,n,\ell} \lambda^2_{n \ell} \frac{
f(\xi_{n, \kv}) \! -\! f(-\xi_{\ell, -\kv+\qv})}{\xi_{n, \kv} \! + \! \xi_{\ell,-\kv+\qv}},
\label{chi}
\end{equation}
where $\xi_{n,\kv}=-2 t_n (\cos k_x + \cos k_y) + \gamma_n - \mu$, and $N_s$ is
the number of lattice sites.
The PDW instability in this model arises from the fact that the dispersion minima of the two bands
(which minimize the denominator in Eq.~\ref{chi}) 
differ in momentum by $\Qv=(\pi,\pi)$. By making a particle-hole transformation (followed by a spin rotation)
of the fermions in the lower band, it is easy to see that the interband singlet Cooper pair
maps onto an exciton.
This PDW instability can thus be recognized as the particle-particle analog of the Halperin-Rice exciton condensation instability in indirect bandgap semiconductors~\cite{HalperinRice}.
The PDW state appears
when strong interband interactions can overcome the insulating band gap. This is most natural in
circumstances where the band insulator and the pairing terms derive from the same microscopic
interactions, such as pairing induced by superexchange interactions in
a spin density wave state
as in the cuprate and pnictide superconductors.
The superfluid to PDW
transition is first-order for this model and the PDW is not accompanied by a charge modulation since $\Qv=(\pi,\pi)$.
More generally, there will be an accompanying $2\Qv$ charge
modulation as well as an
intervening supersolid state.

{\it\bf Experimental signatures. ---}
A direct way to probe for the PDW in solids is a spatially resolved Josephson tunneling experiment \cite{dynes} designed to look for order parameter modulations at wave vector $\Qv$. An indirect signature would be the induced charge
modulation at wave vector $2\Qv$ which one can detect via X-ray scattering \cite{abbamonte}. In cold atom systems,
noise correlations between different spin species \cite{altman} can be used to directly probe nonzero momentum Cooper pairs as has also been proposed for FFLO states \cite{luscher}. An indirect signature would be induced density
modulations at wave vector $2\Qv$ which can be seen from the molecular momentum distribution.

Fluctuations can reduce the PDW wavevector $|\Qv|$ and broaden the momentum distribution peaks (MDP) at $\Qv$. The first effect is not appreciable for $T<E_g$ (band-gap), and a PDW can be observed if $|\Qv|^{-1}$ is smaller than the trap size (or mean-free-path in the presence of disorder), which can be achieved by choosing a suitable scattering length (or a clean material). The second effect is due to the excitation of Goldstone modes with energies $\omega(\qv)$ and momenta $\qv$ away from $\Qv$. Above $T_* \sim \omega(0)-\omega(\Qv)$, where $\omega(\Qv)=0$, the distinct finite-$\Qv$ MDPs will merge into a broad peak at $\qv=0$, and the PDW will revert to a uniform superfluid. For example, at the PDW transition with two fermions per well and $a_L/a=-3.33$ we find $E_g \approx 6.67 E_r \approx 4.8 \mu\textrm{K}$ and $T_* \approx 0.28E_r  \approx 200 \textrm{nK}$ for the circumstances in Ref.\cite{Ketterle2}. In order to maintain phase coherence we must be at temperatures well below the energy scale of the lowest band width, which sets the superfluid stiffness; this leads to an estimated $T_c^0 \sim 1.2 \mu\textrm{K}$. The PDW stability can also be enhanced by going to larger filling factors. We conclude that an atomic PDW is within experimental reach using evaporative cooling techniques to ensure $T \ll T_*, T_c^0$.


{\bf Acknowledgements:}  Some of the numerical calculations were performed on
Rice University computing clusters. PN was supported by W.M. Keck Program in
Quantum Materials. We acknowledge support from NSERC of Canada (AAB and AP) and the
Sloan Foundation (AP).


\begin{thebibliography}{99}
\bibitem{calculation}
These calculations involve numerical diagonalization of the single-particle Hamiltonian with the potential (\ref{Pot}). The lowest $9^3$ bands were treated exactly, the contribution of higher bands was estimated analytically.
\end{thebibliography}
\end{document}